\def\cross{\bds\times}

\def\cross{\bds\times}

\def\rfr#1{Equation\,(\ref{#1})}
\def\rfrs#1#2{Equations\,(\ref{#1})--(\ref{#2})}

\def\virg#1{``#1"}

\def\eqi{\begin{equation}}
\def\eqf{\end{equation}}
\def\eqia{\begin{eqnarray}}
\def\eqfa{\end{eqnarray}}

\def\rp#1#2{\frac{#1}{#2}}
\def\lb#1{\label{#1}}

\def\bds#1{\boldsymbol{#1}}
\def\ton#1{\left(#1\right)}
\def\qua#1{\left[#1\right]}
\def\grf#1{\left\{#1\right\}}

\documentclass[onecolumn]{aastex}

\usepackage{morefloats}
\usepackage[title]{appendix}
\usepackage{textcomp}
\usepackage{booktabs}
\usepackage[table,xcdraw]{xcolor}
\usepackage{multirow}
\usepackage{rotating,tabularx}
\usepackage{float}
\usepackage{enumerate}
\usepackage{rotating}
\usepackage[polutonikogreek,english]{babel}
\usepackage{amsmath,starfont,textgreek,w-greek,wasysym}
\usepackage[flushleft]{threeparttable}
\usepackage{amsthm}
\usepackage{amscd,lineno}
\usepackage{amssymb,dsfont}
\usepackage{graphicx,epsfig}
\usepackage{txfonts}
\usepackage{xr-hyper}
\usepackage{hyperref}
\RequirePackage{color}

\newcommand{\emaila}{lorenzo.iorio@libero.it}

\allowdisplaybreaks[1]

\begin{document}

\title{The Lense-Thirring effect on the Galilean moons of Jupiter}

\shortauthors{L. Iorio}

\author{Lorenzo Iorio\altaffilmark{1} }
\affil{Ministero dell' Istruzione e del Merito
\\ Viale Unit\`{a} di Italia 68, I-70125, Bari (BA),
Italy}

\email{\emaila}

\begin{abstract}
The perspectives  of detecting the general relativistic \textcolor{black}{gravitomagnetic} Lense-Thirring effect on the orbits of the Galilean moons of Jupiter induced by the angular momentum ${\boldsymbol{S}}$ of the latter  are preliminarily investigated. Numerical integrations over one century show that the expected gravitomagnetic signatures of \textcolor{black}{the directly observable right ascension $\alpha$ and declination $\delta$ of the satellites} are as large as tens of arcseconds for Io, while for Callisto they drop to the $\simeq 0.2\,\mathrm{arcseconds}$ level. Major competing effects due to the mismodeling in the zonal multipoles $J_\ell,\,\ell=2,\,3,\,4,\,\ldots$ of the Jovian non-spherically symmetric gravity field and in the Jupiter's spin axis ${\boldsymbol{\hat{k}}}$ should have a limited impact, especially in view of the future improvements in determining such parameters expected after the completion of the ongoing Juno mission in the next few years. \textcolor{black}{On the other hand, the masses of the satellites, responsible of their mutual $N-$body perturbations, should be known better than now. Such a task should be accomplished with the future JUICE and Clipper missions to the Jovian system}. Present-day accuracy in knowing the orbits of \textcolor{black}{the Jovian Galilean satellites} is of the order of 10 milliarcseconds, to be likely further improved \textcolor{black}{thanks to the ongoing re-reduction of old photographic plates}. This suggests that\textcolor{black}{, in the next future,} the Lense-Thirring effect in the main Jovian system of moons might be detectable with dedicated data reductions in which the gravitomagnetic field is explicitly modeled and solved-for.
\end{abstract}

%{
%\textit{Unified Astronomy Thesaurus concepts}:\,Exoplanets\,(498); General relativity\,(641)
%}

%\keywords{General relativity and gravitation; Experimental studies of gravity;  Experimental tests of gravitational theories;  Extrasolar planetary systems}
\section{Introduction}
The linearized equations of the weak-field and slow-motion approximation of the General Theory of Relativity (GTR) \citep{2017grav.book.....M} formally resemble those of the Maxwellian electromagnetism giving rise to the so-called \virg{gravitoelectromagnetic} paradigm \citep{1958NCim...10..318C,Thorne86,1986hmac.book..103T,1988nznf.conf..573T,1991AmJPh..59..421H,
1992AnPhy.215....1J,2001rfg..conf..121M,2001rsgc.book.....R,Mash07,2008PhRvD..78b4021C,
2014GReGr..46.1792C,2021Univ....7..388C,2021Univ....7..451R}. General relativistic gravitoelectromagnetism has, actually, nothing to do with electric charges and currents, implying certain purely gravitational phenomena about orbiting test particles, precessing gyroscopes, moving clocks and atoms, and propagating electromagnetic waves \citep{1977PhRvD..15.2047B,1986SvPhU..29..215D,2002EL.....60..167T,2002NCimB.117..743R,2004GReGr..36.2223S,2009SSRv..148...37S,2009SSRv..148..105S}.
In particular, matter-energy currents produce a so-called \virg{gravitomagnetic} component of the gravitational field accounted for by the off-diagonal components $g_{0i},\,i=1,\,2,\,3$ of the spacetime metric tensor $g_{\upmu\upnu},\,\upmu,\,\upnu=0,\,1,\,2,\,3$. To the first post-Newtonian (1pN) order, in the case of an isolated, slowly spinning body, the source of its gravitomagnetic field is its spin angular momentum $\bds S$ which, among other things, gives rise a non-central, Lorentz-like acceleration on an orbiting test particle \citep{iers10}. It induces secular precessions of the orbit of the latter  \citep{Sof89,1991ercm.book.....B,SoffelHan19}, collectively known as Lense-Thirring (LT) effect \citep{1918PhyZ...19..156L,1984GReGr..16..711M}, although it should be more correctly named as Einstein-Thirring-Lense effect, according to recent studies \citep{2007GReGr..39.1735P,2008mgm..conf.2456P,Pfister2012,Pfister2014}.

General relativistic gravitomagnetism is believed to have a major role in observed complex  processes near spinning black holes involving accretion disks and relativistic jets \citep{1975ApJ...195L..65B,1978Natur.275..516R,1982MNRAS.198..345M,1984ARA&A..22..471R,1988nznf.conf..573T,1999ApJ...525..909A,
2009MNRAS.397L.101I,2009SSRv..148..105S,2013ApJ...778..165V,2016MNRAS.455.1946F}
%\citep{1975ApJ...195L..65B,1978Natur.275..516R,1982MNRAS.198..345M,1984ARA&A..22..471R,1988nznf.conf..573T}.
The hypothesized Penrose Process \citep{2002GReGr..34.1141P,1971NPhS..229..177P,2021Univ....7..416S}, the Blandford-Znajek effect \citep{1977MNRAS.179..433B} and superradiance \citep{1971JETPL..14..180Z} are due to the gravitomagnetic field of a rotating black hole as well; for a review, see, e.g., \citet{2015CQGra..32l4006T} and references. Despite the theoretically predicted large size of the gravitomagnetic effects involved,  the relatively poor knowledge of the parameters characterizing  such astrophysical scenarios and the related uncertainties generally prevent from designing and performing clean and unambiguous tests of general relativity there.
%Thus, it is important to  look towards  better known--although more mundane--systems like, e.g., our solar system even if the expected LT signatures are %far smaller in order to perform, in principle, more trustworthy tests giving better confidence on the existence of gravitomagnetism as predicted by GTR.

In this respect, some years ago, a gravitomagnetic feature was detected in the field of the spinning Earth with the dedicated GP-B spacecraft-based mission \citep{Varenna74} which measured the Pugh-Schiff precessions \citep{Pugh59,Schiff60} of the axes of four gyroscopes carried onboard to an accuracy of the order of 19 per cent  \citep{2011PhRvL.106v1101E}, despite its originally expected level was about $1$ per cent \citep{2001LNP...562...52E}. Somewhat controversial attempts to measure the orbital LT precessions with the Earth's artificial satellites of the LAGEOS type \citep{2019JGeod..93.2181P} and the Satellite Laser Ranging (SLR) technique \citep{SLR11} are currently ongoing \citep{2013NuPhS.243..180C,2013CEJPh..11..531R,2013AcAau..91..141I}; for other proposed tests with natural and artificial bodies in the solar system, see \citet{2011Ap&SS.331..351I} and references therein. A tight binary system made of a white dwarf and a pulsar was recently used to claim a successful detection of the gravitomagnetic orbital precession of the inclination \citep{2020Sci...367..577V}; weak points of such a test were highlighted in \citet{2020MNRAS.495.2777I}. The double pulsar PSR J0737-3039A/B \citep{2003Natur.426..531B,2004Sci...303.1153L} is another astrophysical laboratory of interest to attempt to measure its gravitomagnetic periastron precession in the next future  \citep{Kehletal017,2020MNRAS.497.3118H}.

Here, the possibility of extracting the LT signatures from long data records of accurate observations \citep{2019JAHH...22...78A} of the Galilean moons  of Jupiter \citep{Schlosser1991} is examined. Future approved missions like Jupiter Icy Moons Explorer\footnote{It was launched on \textcolor{black}{14} April 2023. See the mission's webpage \url{https://www.esa.int/Science\textunderscore Exploration/Space\textunderscore Science/Juice} on the Internet.} (JUICE) \citep{2013P&SS...78....1G}  and Clipper\footnote{Its launch is scheduled for October 2024. See the mission's website \url{https://www.jpl.nasa.gov/missions/europa-clipper} on the Internet.} \citep{2022EGUGA..24.6052K} should further increase the accuracy in determining their \textcolor{black}{masses and} orbits \textcolor{black}{\citep{2020P&SS..18704902C,2021AeMiS.100..195M,2022PSJ.....3..199C,2022P&SS..21905531F}} which, currently, should be of the order of\footnote{R.~A. Jacobson, personal communication to L. Iorio, March 2023.} $\simeq 10\,\mathrm{milliarcseconds\,(mas)}$; see also \citet{2019JAHH...22...78A}. More precisely, the present goal is primarily pointing out that the theoretically predicted LT effect for the main Jovian satellites should fall within the measurability domain since it is much larger than the current astrometric accuracy level for them. \textcolor{black}{A preliminary assessment of the major source of systematic biases is provided as well.} It should be remarked that it would be incorrect to straightforwardly compare theoretical calculation of such an effect to residuals currently existing in the literature obtained without including the LT acceleration itself in the models used to produce them \citep{2023arXiv230301821F}. Actually, the LT acceleration is nowadays modeled\footnote{R.~A. Jacobson, private communication to L.~Iorio, March 2023.} in the dynamics of the Galilean moons of Jupiter (and of \textcolor{black}{the Juno spacecraft} as well, \textcolor{black}{currently exploring the Jovian system}), but, so far, no dedicated data reductions aimed to test it were performed. \textcolor{black}{Furthermore, it is hoped that the present work will motivate dedicated investigations of the JUICE and Clipper teams who, hopefully, may explicitly model the Jovian gravitomagnetic field in their covariance analyses aimed at assessing the level of improvement in the ephemerides of the Galilean moons which will be achievable with such missions in connection with the proposed test.}

The paper is organized as follows. In Section\,\ref{sect1}, the LT signatures of some observable quantities of Io, Europa, Ganymede and Callisto are numerically calculated over one century and are plotted versus time. Section\,\ref{sect2} is devoted to the numerically integrated time series of the same observables due to the mismodeling in the  multipoles of the Jovian gravity field and the uncertainties in the position in space of the spin axis of Jupiter. \textcolor{black}{Section\,\ref{sect3} deals with the mutual $N-$body perturbations of one satellite on each other.} Section\,\ref{sect4} summarizes the findings of the paper and offers its conclusions.
\section{The Lense-Thirring signatures}
\begin{figure}
\centering
\begin{tabular}{cc}
\includegraphics[width = 7 cm]{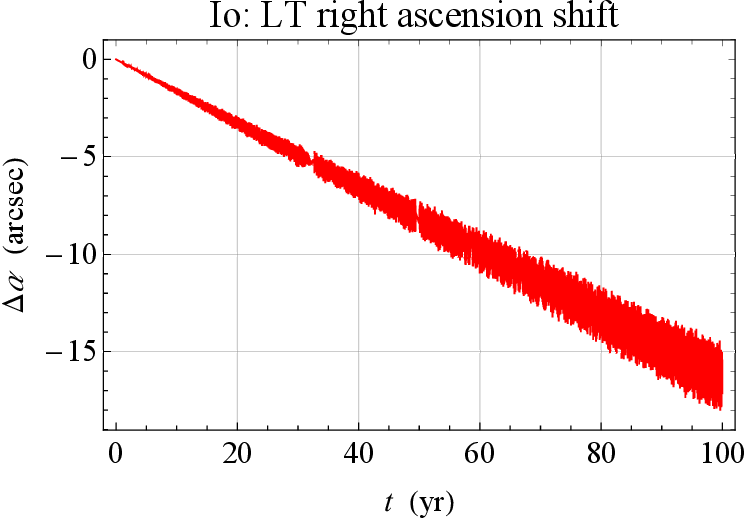}& \includegraphics[width = 7 cm]{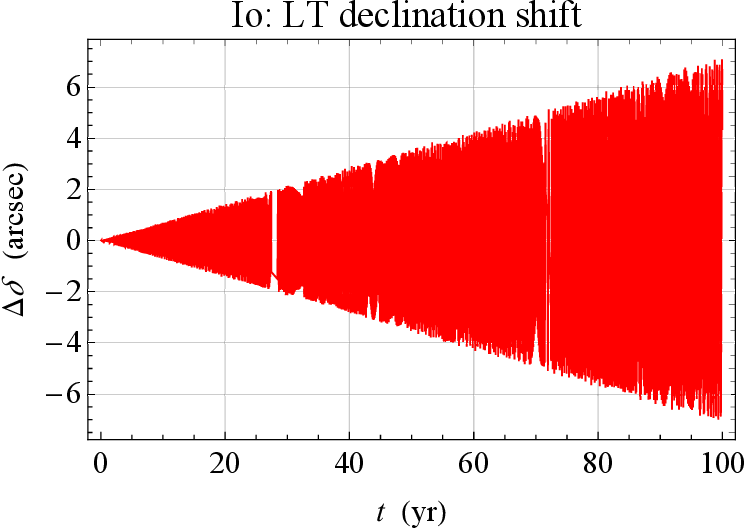}\\
\includegraphics[width = 7 cm]{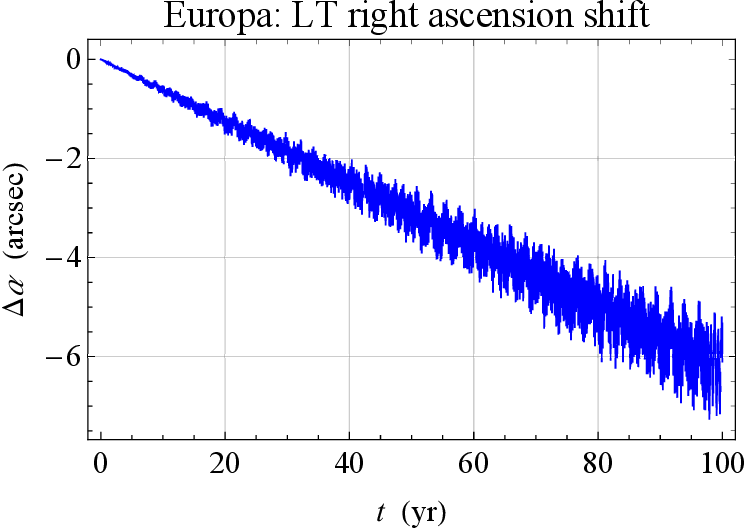}& \includegraphics[width = 7 cm]{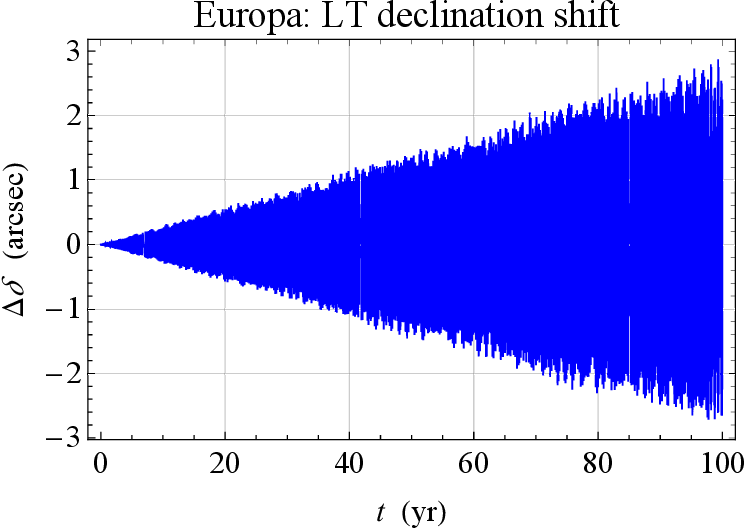}\\
\includegraphics[width = 7 cm]{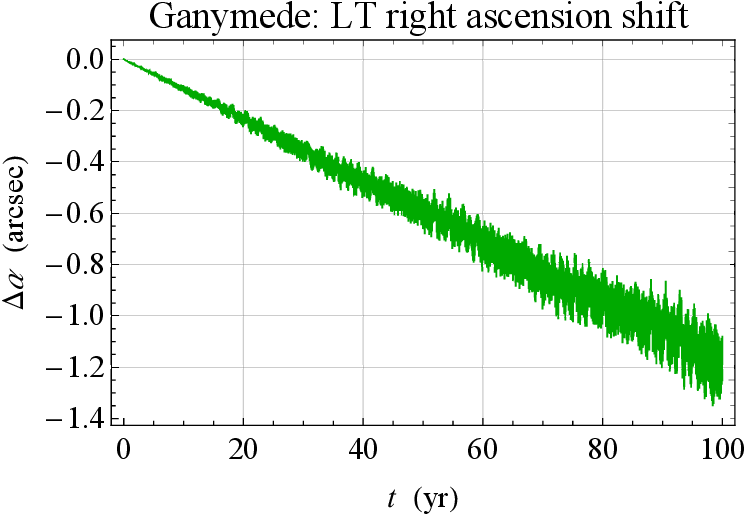}& \includegraphics[width = 7 cm]{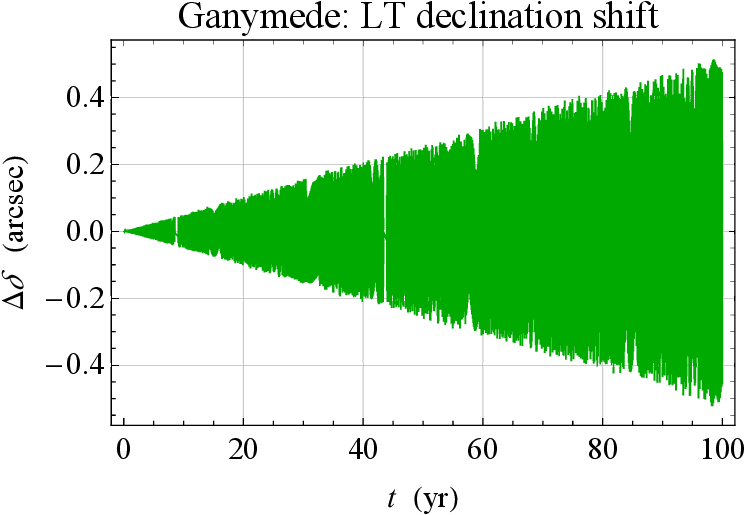}\\
\includegraphics[width = 7 cm]{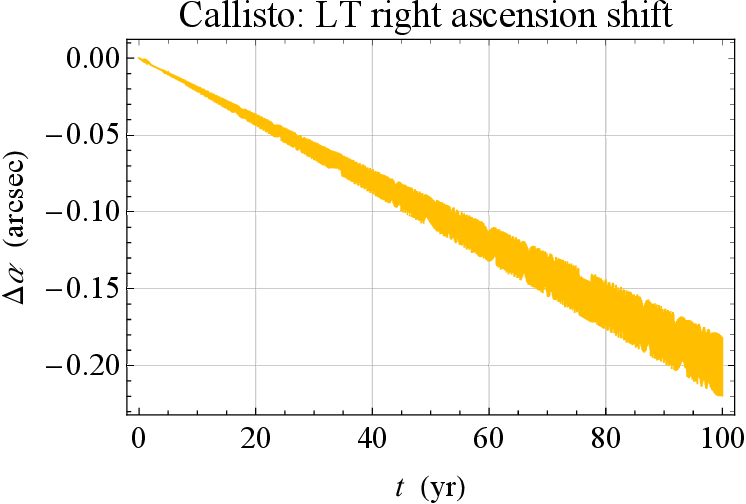}& \includegraphics[width = 7 cm]{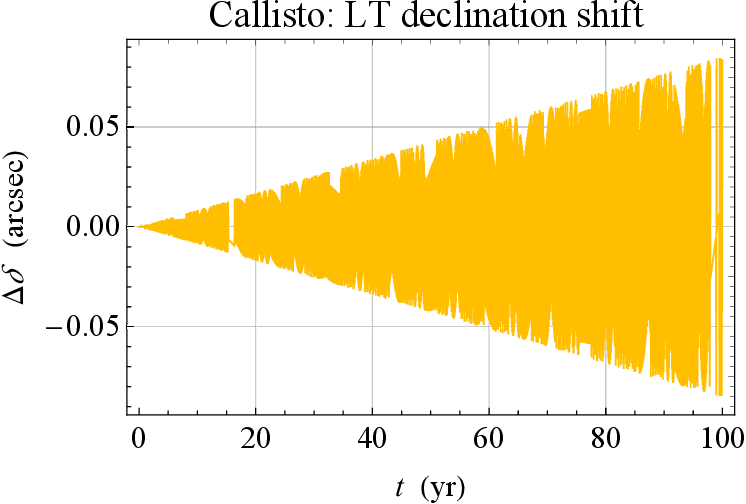}\\
\end{tabular}
\caption{
\textcolor{black}{Numerically produced time series of the nominal LT shifts $\Delta\alpha$ and $\Delta\delta$, in arcseconds (arcsec), of the RA and the DEC of the Galilean moons of Jupiter with respect to the ICRF.}
}\label{fig1}
\end{figure}

\lb{sect1}
\textcolor{black}{
Figure\,\ref{fig1} displays the nominal centennial LT signatures of the astrometrically measurable right ascension (RA) $\alpha$ and  declination (DEC) $\delta$ of Io, Europa, Ganymede, and Callisto. Their LT time series $\Delta\alpha\ton{t}$ and $\Delta\delta\ton{t}$ were obtained as follows.
Their equations of motion, with and without the LT acceleration\textcolor{black}{\footnote{\textcolor{black}{In \rfr{ALT}, $G$ is the Newtonian constant of gravitation, $c$ is the speed of light in vacuum, and $\bds r$ and $\bds v$ are the position and velocity vectors of the test particle with respect to the spinning central body, respectively.}}} \citep{iers10}
\eqi
{\bds{A}}_\mathrm{LT} =\rp{2 G S}{c^2 r^3}\qua{\rp{3}{r^2}\ton{\bds{\hat{k}}\bds\cdot\bds r}\ton{\bds r\bds\cross\bds v} -\ton{\bds{\hat{k}}\bds\cross\bds v}}\lb{ALT}
\eqf
calculated with \citep{2003AJ....126.2687S}
}
\eqi
S \simeq 6.9\times 10^{38}\,\mathrm{kg\,m^2\,s^{-1}},
\eqf
and
\begin{align}
{\hat{k}}_x \lb{kx} & = \cos\alpha_{\jupiter}\,\cos\delta_{\jupiter}, \\ \nonumber \\
{\hat{k}}_y \lb{ky} & = \sin\alpha_{\jupiter}\,\cos\delta_{\jupiter}, \\ \nonumber \\
{\hat{k}}_z \lb{kz} & = \sin\delta_{\jupiter},
\end{align}
with
\citep{2020GeoRL..4786572D}
\begin{align}
\alpha_{\jupiter} \lb{RAJ} & = 268.05656^\circ, \\ \nonumber \\
\delta_{\jupiter} \lb{DECJ} & = 64.49530^\circ,
\end{align}
\textcolor{black}{for the orientation of the Jovian spin axis ${\bds{\hat{k}}}$,
were simultaneously integrated over one century in a coordinate system  using rectangular Cartesian coordinates centered in the Jovian system barycenter and having the mean Earth's equator at the reference epoch J2000 as reference $\grf{x,\,y}$ plane; \rfrs{RAJ}{DECJ} refer just to it.
Both the integrations shared  the same initial conditions, referred to the International Celestial Reference Frame (ICRF) at some epoch and retrieved from the HORIZONS WEB interface maintained by the NASA Jet Propulsion Laboratory (JPL). Moreover, the suite of dynamical models, including for each satellite the Newtonian inverse-square law accelerations due to Jupiter and the other three moons, the Jovian gravitational multipoles up to degree $\ell=6$, and the general relativistic Schwarschild-type gravitoelectric acceleration due to Jupiter, is the same for both the runs. For each integration, a time series was numerically computed for $\alpha$ and $\delta$. The differences $\Delta\alpha\ton{t}$ and $\Delta\delta\ton{t}$ between the times series with and without the LT effect were, then, calculated and plotted versus time.}
\clearpage
\textcolor{black}{
It turns out that, for Io, the LT centennial shift on the RA looks like a negative linear trend as large as $\simeq 20\,\mathrm{arcseconds}$ (arcsec), while the LT signal of its DEC has a peak-to-peak amplitude of $\simeq 14\,\mathrm{arcsec}$.
The patterns of the LT signatures of Europa, Ganymede and Callisto are the same; the sizes of the linear trends $\Delta\alpha$ of their RA amount to approximately $6,\,1.3$, and $0.20$ arcsec, respectively, while the peak-to-peak amplitudes of the DEC shifts $\Delta\delta$ are about $6,\,1$ and $0.2$ arcsec, respectively.
}

\textcolor{black}{
In recent years, an effort of reducing old observations using today's accuracy has been undertaken. Indeed, at the beginning
of the current century, it became possible to scan and digitize old photographic plates with modern scanners accurate to a few nanometers, and to reduce them using the new accurate catalogues of reference stars \citep{2019JAHH...22...78A}. As an example, Table 15 of Arlot \citep{2019JAHH...22...78A} displays the accuracy of newly digitized and re-reduced photographic plates spanning a 4 yr long time interval (1986-1990) in terms of their dispersion $\sigma$ and mean observed-minus-calculated (O-C) differences.
}

\textcolor{black}{
Over 4 yr, the expected LT RA trend of Io reaches the $0.8$ arcsec level, while the peak-to-peak amplitude of its LT DEC signal is about $0.6$ arcsec. Such figures are to be compared with those quoted in Table 15 of \citet{2019JAHH...22...78A} which, in the case of Io, are as little as $0.027$ ($\sigma$) and $0.002$ arcsec (O-C) for the RA, and $0.040$ ($\sigma$) and $0.010$ arcsec (O-C) for the DEC. Thus, at least in principle, a successful LT detection at about the percent level, or, perhaps, even better, might cautiously be possible by exploiting the astrometric observations of Io.
}

\textcolor{black}{
As far as Europa is concerned, its predicted LT signatures over 4 yr are about $0.3$ (RA trend) and $0.2$ arcsec (peak-to-peak DEC amplitude), respectively. For the second Galilean moon of Jupiter, Table 15 of \citet{2019JAHH...22...78A} returns $0.023$ ($\sigma$) and $0.0005$ arcsec (O-C) for the RA, and $0.039$ ($\sigma$) and $0.007$ arcsec (O-C) for the DEC. Such errors range from some percent to about 19 percent of the LT effect.
}

\textcolor{black}{
Table 15 of \citet{2019JAHH...22...78A} yields $0.020$ ($\sigma$) and $0.002$ arcsec (O-C) for the RA, and $0.039$ ($\sigma$) and $0.002$ arcsec (O-C) for the DEC of Ganymede. The predicted LT effects for the third Galilean moon of Jupiter are $0.05$ (RA linear trend) arcsec and $0.04$ arcsec (peak-to-peak DEC amplitude) over 4 yr. In the most favorable case, the observational uncertainties are at the percent level of the LT signals, while in the worst case, they are as large as $\simeq 40-97\%$ of them.
}

\textcolor{black}{
Finally, the expected LT shifts of Callisto are  $0.009$ arcsec (RA linear trend) arcsec and $0.006$ arcsec (peak-to-peak DEC amplitude) over 4 yr. The corresponding observational errors, according to Table 15 of \citet{2019JAHH...22...78A}, are $0.019$ ($\sigma$) and $0.0002$ arcsec (O-C) for the RA, and $0.022$ ($\sigma$) and $0.004$ arcsec (O-C) for the DEC. Apart from the RA O-C, which is as little as $\simeq 2\%$ of the LT shift, the other errors are larger than, or of the same order of magnitude of the relativistic targets.
}

\textcolor{black}{
The preliminary comparisons just presented suggests that the Jovian LT effect on the Galilean moons falls within the measurability domain to a level of accuracy which, at least for some of them, should be deemed as competitive with the GP-B experiment, which is the only currently undisputed existing test of the gravitomagnetic field.
This conclusion should further strengthen as the re-reduction  of the old photographic plates will go on by processing longer data records.
}
\section{The impact of the  zonal harmonics of the Jovian multipolar gravity field}\lb{sect2}
A potentially major source of systematic error in recovering the LT effect by means of the Galilean satellites of Jupiter is represented by the accuracy with which the multipolar expansion of the classical part $\Delta U\ton{\bds r}$ of the Jovian gravitational potential $U\ton{\bds r}$, accounting for its departures from spherical symmetry, will be known at the time of an actual data analysis aimed to extract the relativistic signatures of interest. In particular, both the zonal harmomic coefficients $J_{\ell},\,\ell=2,\,3,\,\ldots$, in terms of which $\Delta U$ is parameterized, and the Jupiter's spin axis ${\bds{\hat{k}}}$, entering $\Delta U$ itself, need to be accurately known.
\textcolor{black}{
The correction of degree $\ell$ to the gravitational potential reads
\eqi
\Delta U_\ell\ton{\bds r} = \rp{GM}{r}J_\ell\ton{\rp{R}{r}}^\ell\mathcal{P}_\ell\ton{\bds{\hat{k}}\bds\cdot\bds{\hat{r}}},
\eqf
where $R$ is the equatorial radius of the primary, and $\mathcal{P}_\ell\ton{\cdots}$ is the is the Legendre polynomial
of degree $\ell$ whose argument is the cosine of the angle between $\bds{\hat{k}}$ and the position vector $\bds r$ of the
test particle.
}

\begin{figure}
\centering
\begin{tabular}{cc}
\includegraphics[width = 7 cm]{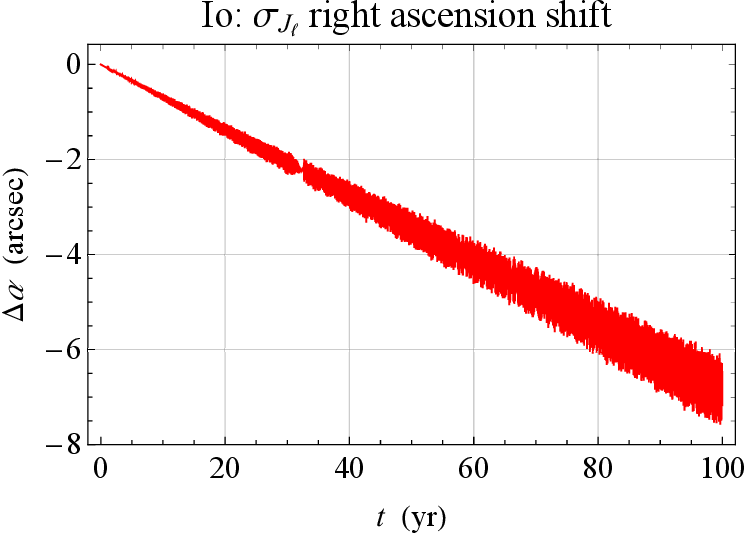}& \includegraphics[width = 7 cm]{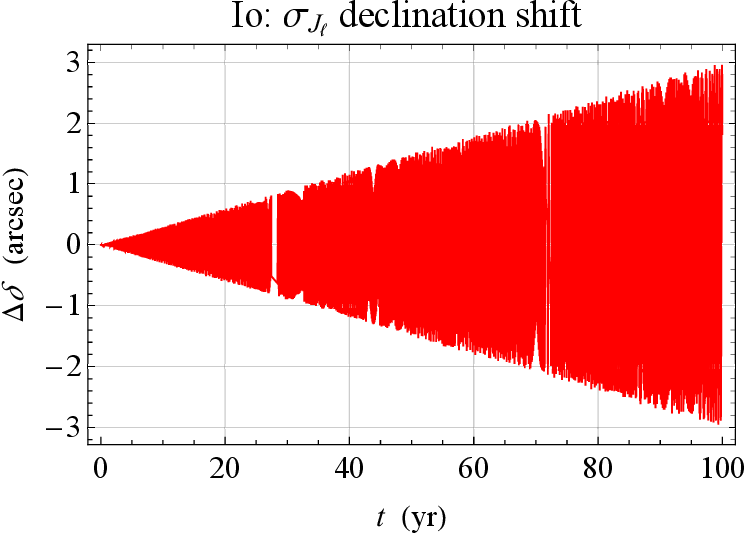}\\
\includegraphics[width = 7 cm]{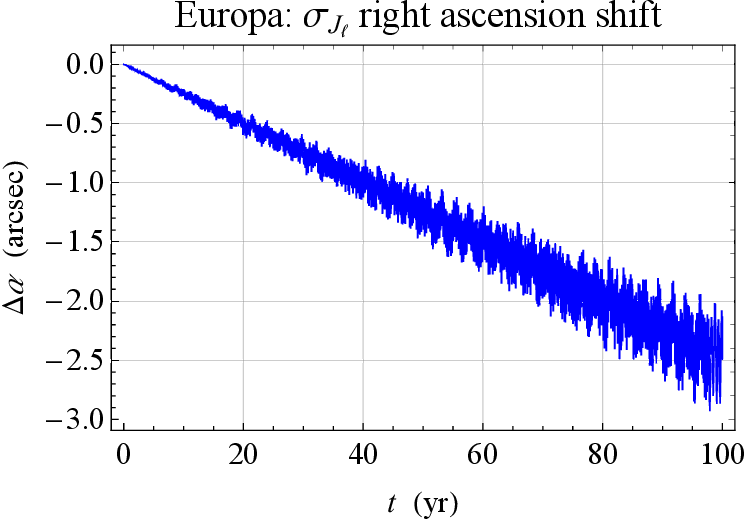}& \includegraphics[width = 7 cm]{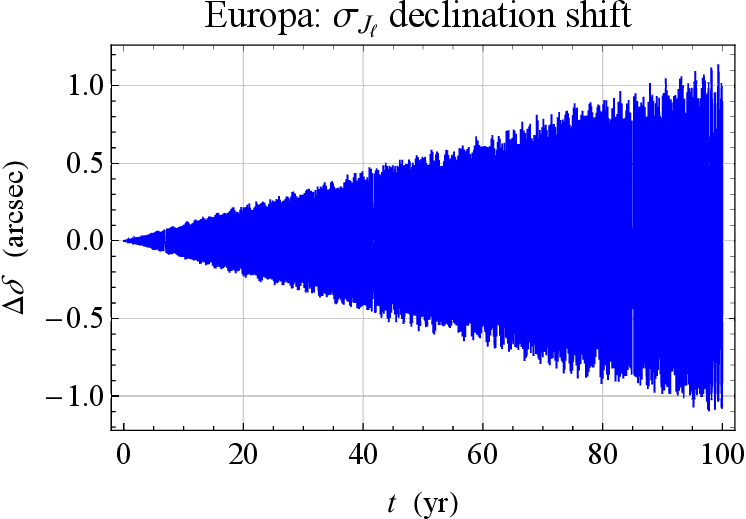}\\
\includegraphics[width = 7 cm]{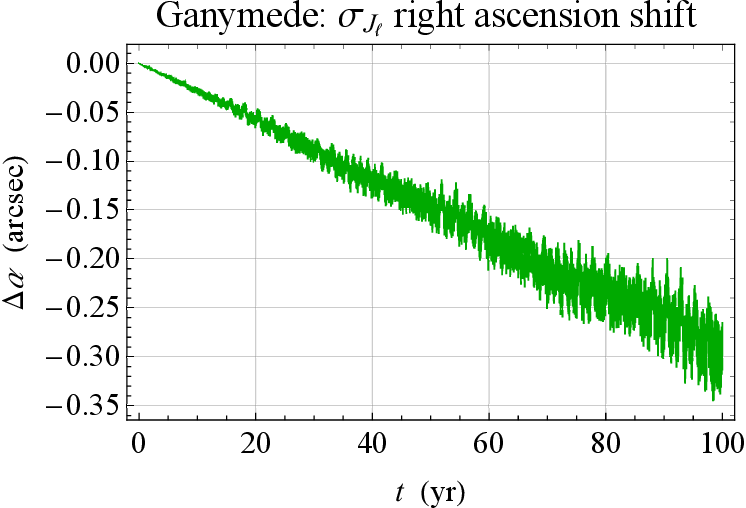}& \includegraphics[width = 7 cm]{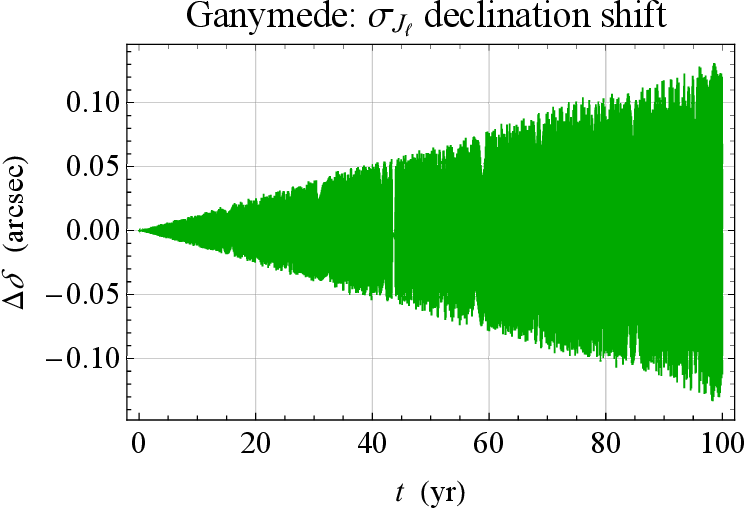}\\
\includegraphics[width = 7 cm]{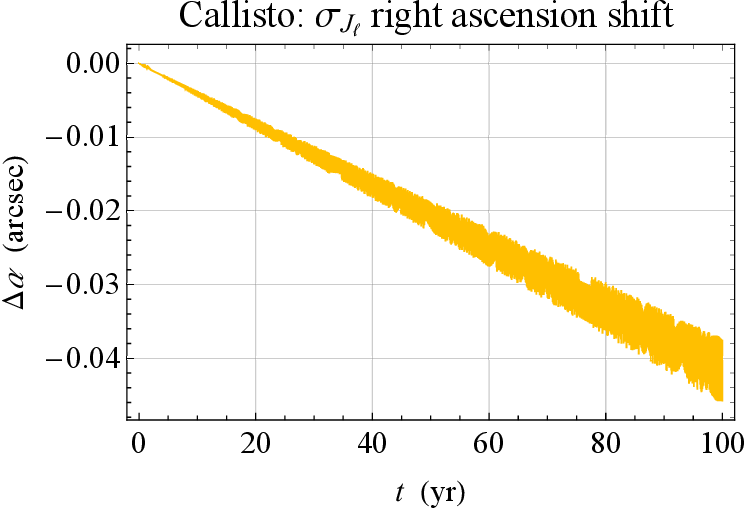}& \includegraphics[width = 7 cm]{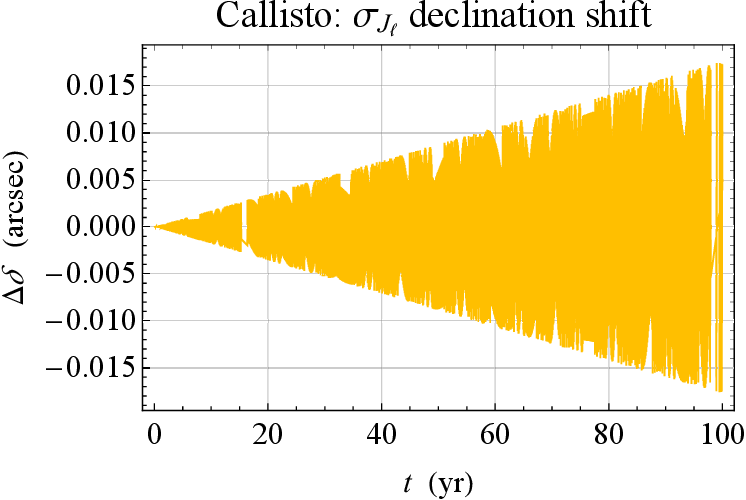}\\
\end{tabular}
\caption{
\textcolor{black}{Numerically produced time series of the first five mismodelled  zonals  shifts $\Delta\alpha$ and $\Delta\delta$, in arcseconds, of the RA and the DEC of the Galilean moons of Jupiter with respect to the ICRF. The uncertainties $\sigma_{J_\ell},\,\ell=2,\,3,\,4,\,5,\,6$  in the Jovian gravity field multipoles were retrieved from \citet{2020GeoRL..4786572D}.}
}\label{fig2}
\end{figure}

\textcolor{black}{In Figure\,\ref{fig2}, the signatures of RA and DEC} due to the current level of mismodeling in $J_\ell,\,\ell=2,\,3,\,\ldots$  up to degree $\ell=6$ are displayed.  The uncertainties $\sigma_{J_\ell},\,\ell=2,\,3,\,4,\,5,\,6$  were retrieved from \citet{2020GeoRL..4786572D}, and are based just on the first half of the currently ongoing Juno mission \citep{juno18}. When all the data record will be available after its completion in the next few years, they will certainly be further improved. The time series shown were built as in Section\,\ref{sect1} for the LT effect; in this case, for a given satellite, one run was performed with $J_\ell + \sigma_{J_\ell}$, while the other integration was made with $J_\ell - \sigma_{J_\ell}$, respectively, all the rest of the dynamical model being equal in both the runs. In particular, the orientation of the Jovian spin axis, entering explicitly the multipole-induced accelerations, was kept fixed to the same value, given by \rfrs{RAJ}{DECJ},
in both the integrations.
\clearpage
\textcolor{black}{It turns out that even the current level of accuracy in knowing the Jupiter's gravity multipoles would be adequate to yield a systematic error smaller than the LT effect. Indeed, for Io, the RA and DEC centennial signatures due to $\sigma_{J_\ell}$ would be of the order of just $\simeq 8\,\mathrm{arcsec}$ (RA linear trend) and $6$ arcsec (DEC peak-to-peak amplitude), i.e. $\simeq 2.5-3$ times smaller than the corresponding LT effects.
For Callisto, the RA and DEC shifts due to $\sigma_{J_\ell}$ amount to $0.04$ arcsec (RA linear trend) and $0.03$ arcsec (DEC peak-to-peak amplitude), i.e., $\simeq 5-6$ times smaller than the LT ones.
}

\begin{figure}
\centering
\begin{tabular}{cc}
\includegraphics[width = 7 cm]{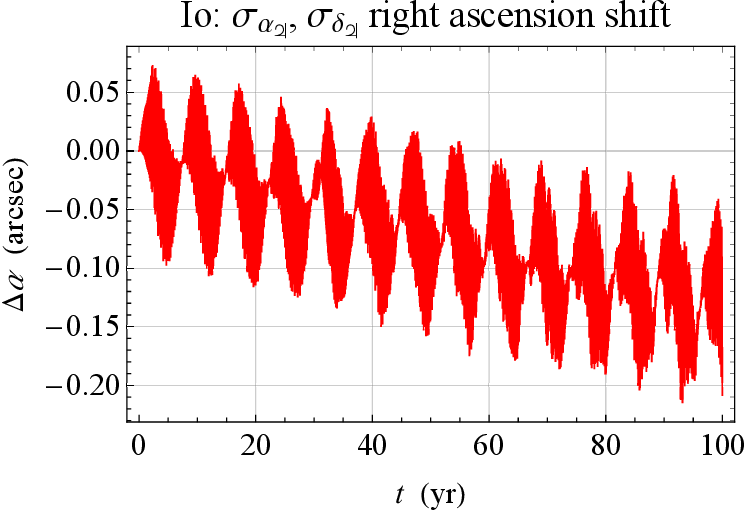}& \includegraphics[width = 7 cm]{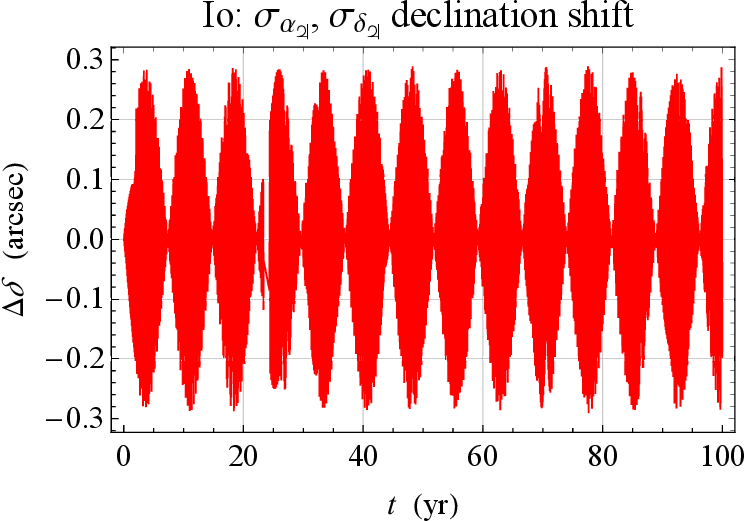}\\
\includegraphics[width = 7 cm]{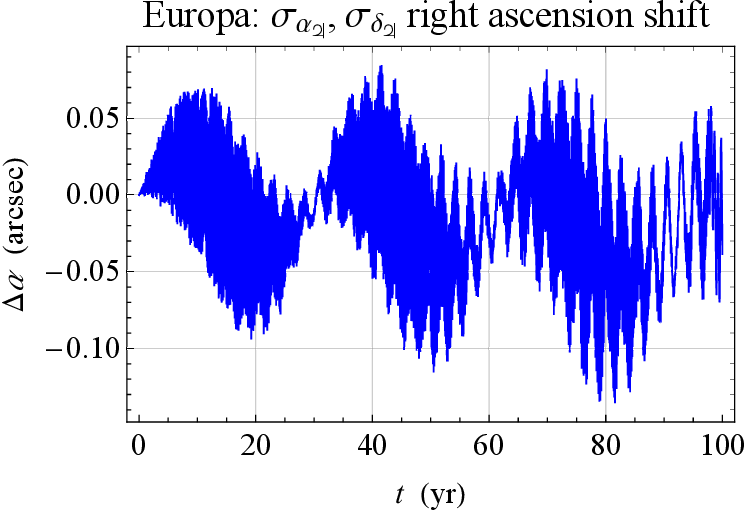}& \includegraphics[width = 7 cm]{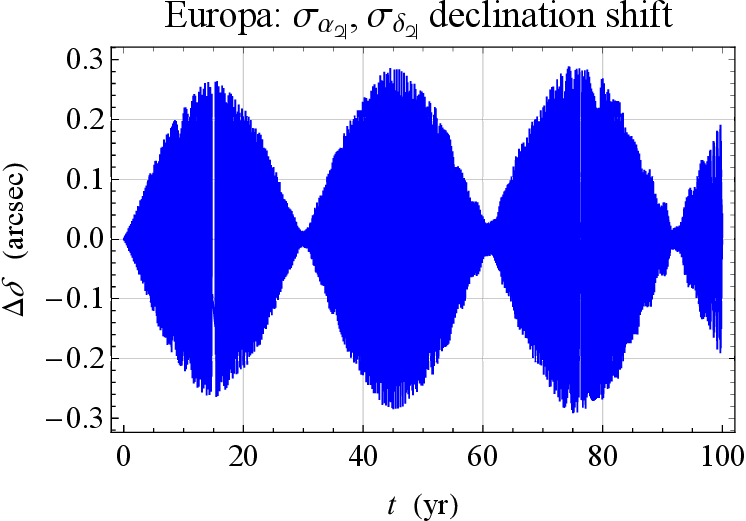}\\
\includegraphics[width = 7 cm]{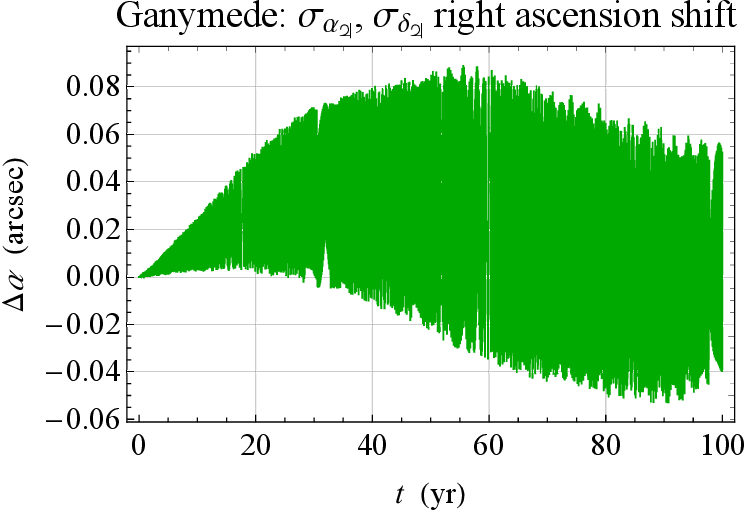}& \includegraphics[width = 7 cm]{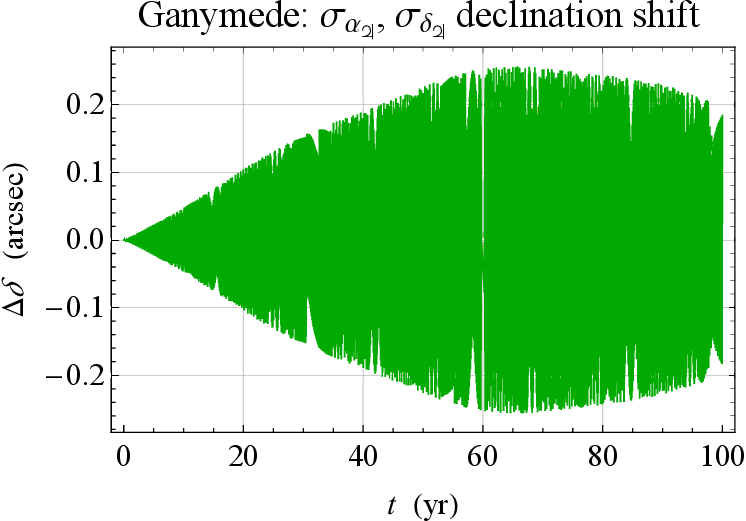}\\
\includegraphics[width = 7 cm]{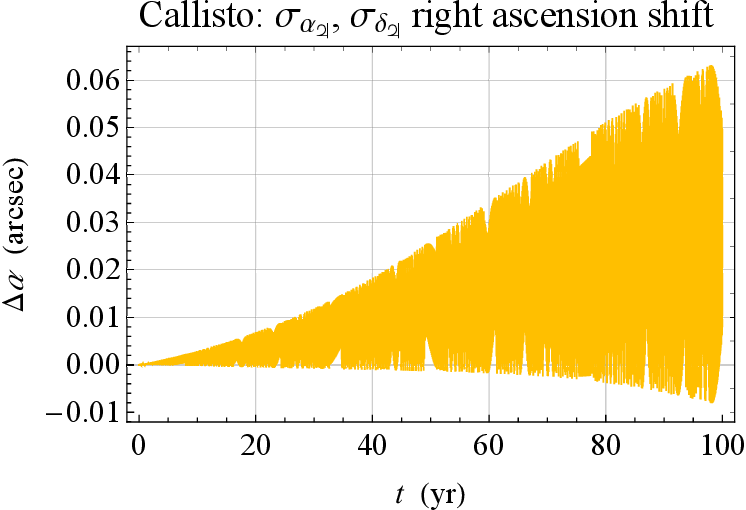}& \includegraphics[width = 7 cm]{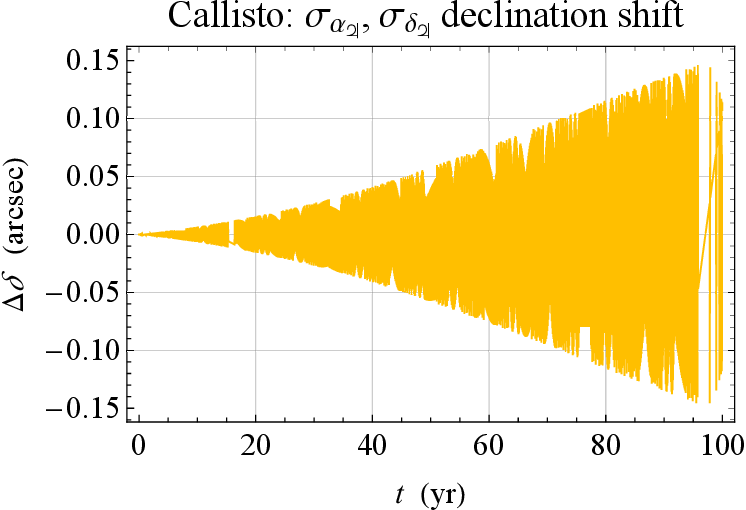}\\
\end{tabular}
\caption{\textcolor{black}{Impact of the current uncertainty in the Jovian pole of rotation through the first five zonal harmonics of the gravity field on the RA and the DEC of the Galilean moons of Jupiter with respect to the ICRF; the resulting shifts $\Delta\alpha$ and $\Delta\delta$ are in arcseconds. The uncertainties in ${\bds{\hat{k}}}$ are as per \rfrs{errra}{errdec}.}
}\label{fig3}
\end{figure}

Another potential major source of systematic error for the extraction of the LT signals is represented by the impact of the uncertainty in the Jupiter's spin axis ${\bds{\hat{k}}}$ parameterized in terms of the right ascension $\alpha_{\jupiter}$ and declination $\delta_{\jupiter}$ of the Jovian pole of rotation as per \rfrs{kx}{kz}.
The RA and DEC of the Jovian spin axis are currently known to the
\begin{align}
\sigma_{\alpha_{\jupiter}} \lb{errra}&\simeq 0.04\,\mathrm{arcsec} = 0.00001^\circ,\\ \nonumber \\
\sigma_{\delta_{\jupiter}}\lb{errdec}&\simeq 0.07\,\mathrm{arcsec} = 0.00002^\circ
\end{align}
level, as it can be inferred from Figure\,3 of \citet{2020GeoRL..4786572D}. It is arguable that also the knowledge of the planet's pole of rotation will be further improved when all the complete record of Juno's data will be available at the end of the mission.
\textcolor{black}{Figure\,\ref{fig3} shows the mismodelled RA and DEC signatures} due to the planetary zonal harmonics up to degree $\ell = 6$ induced by the uncertainty in ${\bds{\hat{k}}}$ according to \rfrs{errra}{errdec}. For each satellite, they were obtained  from the difference between two numerically integrated time series one of which calculated with $\alpha_{\jupiter} + \sigma_{\alpha_{\jupiter}},\,\delta_{\jupiter} + \sigma_{\delta_{\jupiter}}$ and the other one with  $\alpha_{\jupiter} - \sigma_{\alpha_{\jupiter}},\,\delta_{\jupiter} - \sigma_{\delta_{\jupiter}}$, all the rest being equal for both the runs.
\clearpage
It turns out that the current level of uncertainty in the Jupiter's spin axis, to be improved in the ongoing future after the completion of the Juno mission, has a modest impact on the LT signature of Io. Indeed, \rfrs{errra}{errdec} induce mismodelled effects of the order of     \textcolor{black}{$\simeq 0.20$ arcsec (RA trend) and $0.6$ arcsec (peak-to-peak DEC amplitude) for Io; they are about 100 times smaller than the relativistic signals displayed in Figure\,\ref{fig1}.}

\textcolor{black}{
Also for Europa the situation is good since the mismodelled signals are $\simeq 10-100$ times smaller than the relativistic ones in Figure\,\ref{fig1}.
}

\textcolor{black}{
The mismodelled signature of the RA of Ganymede is about one order of magnitude smaller than the corresponding relativistic ones displayed in Figure\,\ref{fig1}. For the DEC, the situation is less favorable since the impact of \rfrs{errra}{errdec} is as large as $\simeq 60\%$ of the LT effect.
}

For Callisto, the mismodelled signals due to \rfrs{errra}{errdec} are \textcolor{black}{$\simeq 0.06$} arcsec (RA trend) and \textcolor{black}{$0.3$} arcsec (peak-to-peak DEC amplitude); according to Figure\,\ref{fig1}, the corresponding nominal LT signal for \textcolor{black}{the RA is $\simeq 4$} times larger, \textcolor{black}{while the LT DEC one is $1.5$ times smaller than that due to the uncertainty in $\bds{\hat{k}}$}.
\section{\textcolor{black}{The impact of the Newtonian $N-$body mutual perturbations}}
\begin{figure}
\centering
\begin{tabular}{cc}
\includegraphics[width = 7 cm]{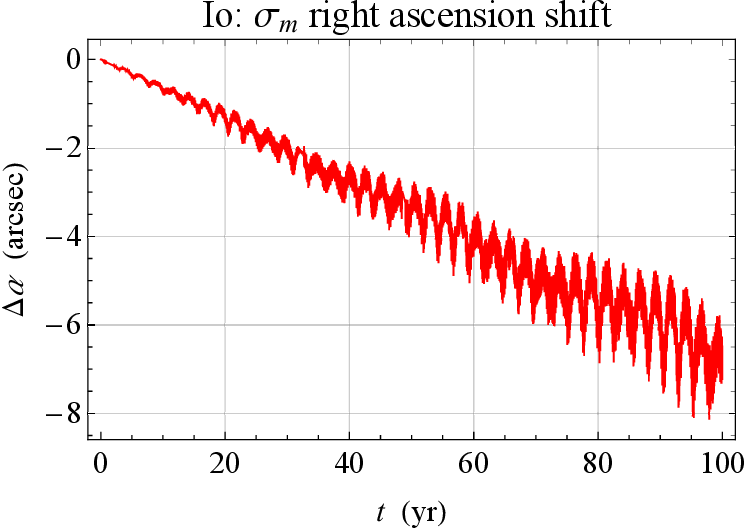}& \includegraphics[width = 7 cm]{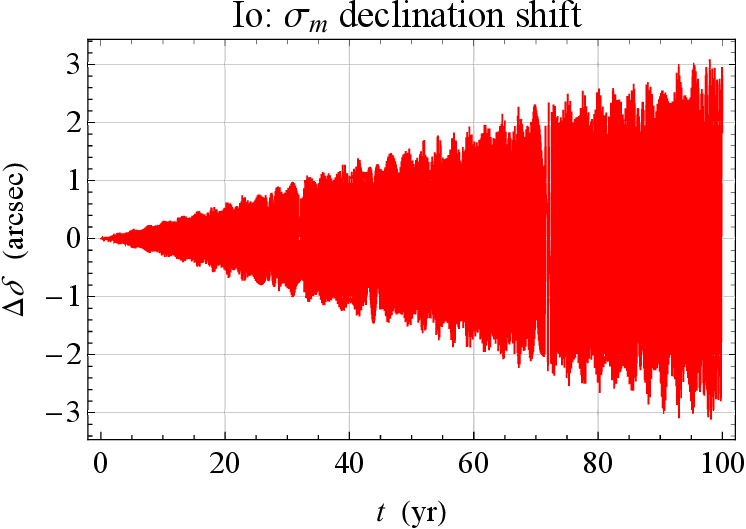}\\
\includegraphics[width = 7 cm]{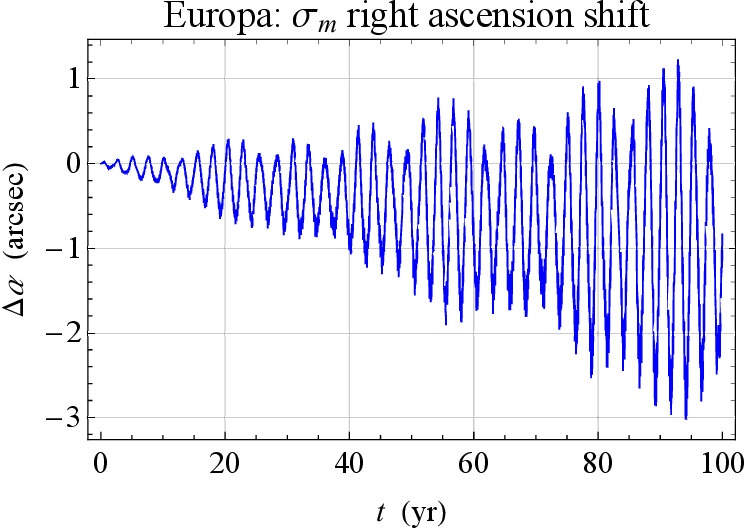}& \includegraphics[width = 7 cm]{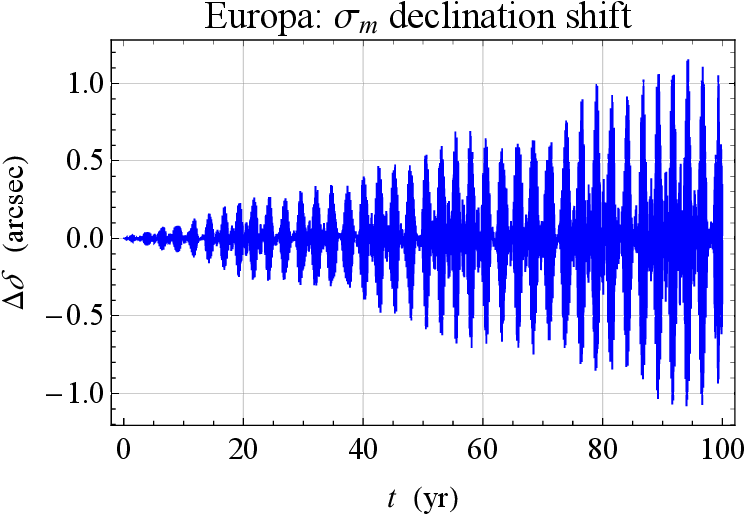}\\
\includegraphics[width = 7 cm]{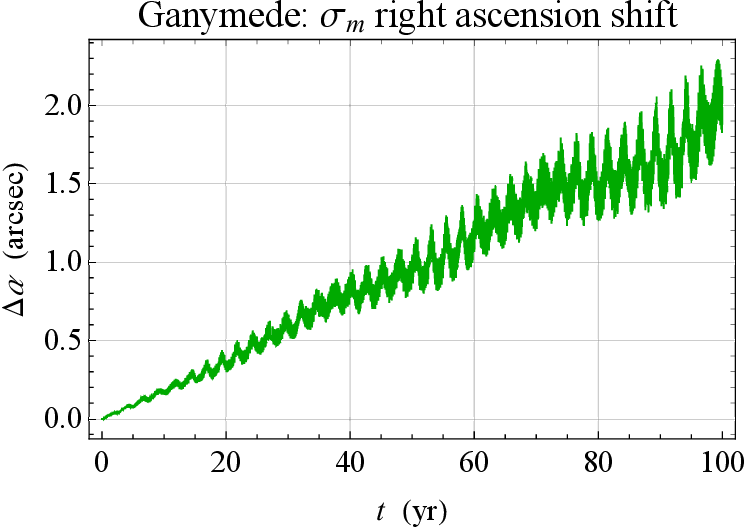}& \includegraphics[width = 7 cm]{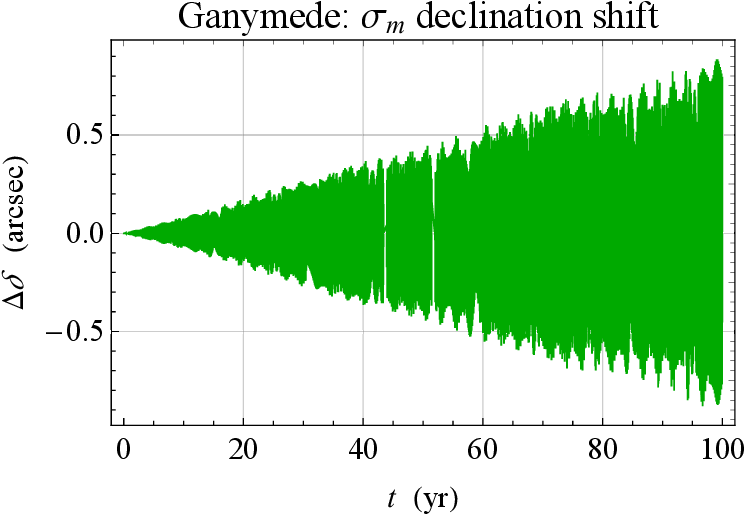}\\
\includegraphics[width = 7 cm]{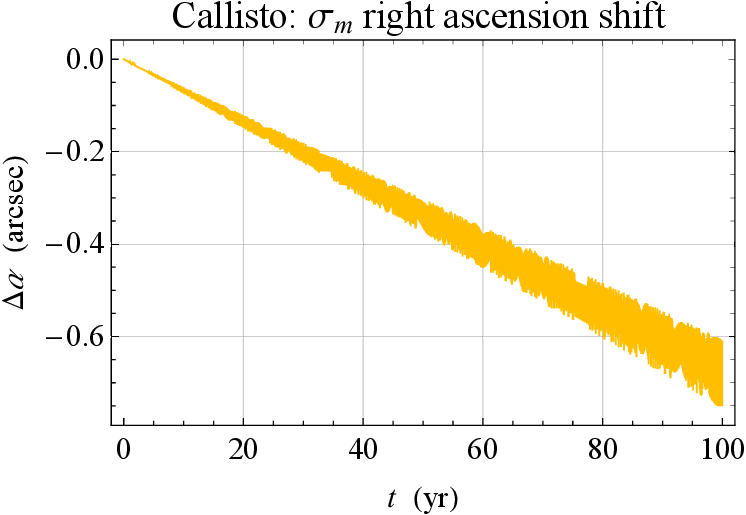}& \includegraphics[width = 7 cm]{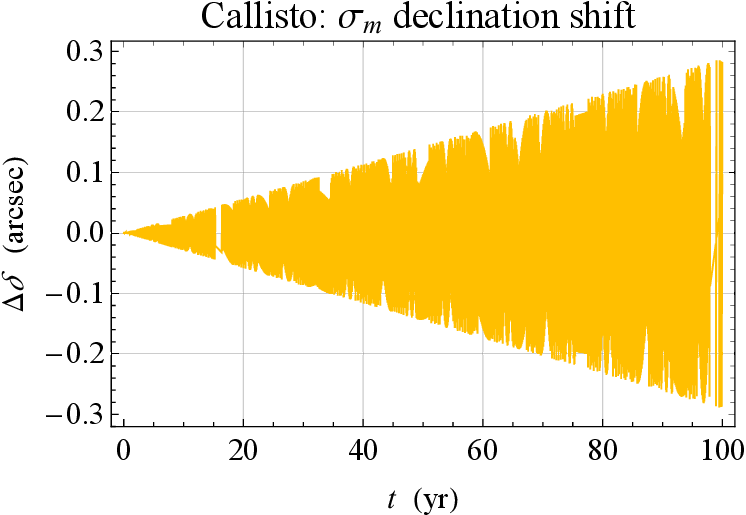}\\
\end{tabular}
\caption{
\textcolor{black}{Numerically produced time series of the mismodelled Newtonian $N-$body shifts $\Delta\alpha$ and $\Delta\delta$, in arcseconds, of the RA and the DEC of the Galilean moons of Jupiter with respect to the ICRF. The uncertainties $\sigma_m$  in the masses of Io, Europa, Ganymede and Callisto were retrieved from  the Planetary Satellite Ephemeris: JUP365 \citep{jup365}, and are available on the Internet at \url{https://ssd.jpl.nasa.gov/sats/phys_par/}.}
}\label{fig4}
\end{figure}

\lb{sect3}
\textcolor{black}{
Another potentially major source of systematic bias is represented by the classical $N-$ body perturbations induced on each satellite by the Newtonian attraction of the other ones. The level of their mismodelling is set by the uncertainty with which the masses $m$ of the Galilean satellites are known.
Figure,\,\ref{fig4} shows the numerically integrated time series $\Delta\alpha$ and $\Delta\delta$ of the RA and the DEC of Io, Europa, Ganymede and Callisto due to the current errors $\sigma_m$ in their masses retrieved on the Internet at \url{https://ssd.jpl.nasa.gov/sats/phys_par/} where the Planetary Satellite Ephemeris: JUP365 \citep{jup365} is quoted as source. They were calculated by subtracting two runs differing only by the values of the masses which are set to $m+\sigma_m$ and $m-\sigma_m$, respectively, all the rest being equal.
}
\clearpage
\textcolor{black}{For Io, perturbed by Europa, Ganymede and Callisto, the RA is affected to the $\simeq 8$ arcsec level over one century, while the peak-to-peak amplitude of its DEC is $6$ arcsec over the same time span. Such figures are smaller than the LT ones of Figure\,\ref{fig1} by $\simeq 2$ times.
}

\textcolor{black}{
The mismodelled signatures of Europa amount to about $4$ arcsec (RA) and $2$ arcsec (DEC), which represent a significative amount of the LT ones in Figure\,\ref{fig1}.
}

\textcolor{black}{
The mismodelled shifts of Ganymede are about $2.5$ arcsec (RA) and $2$ arcsec (DEC), which are $\simeq 2$ times larger than the LT signals displayed in Figure\,\ref{fig1}.
}

\textcolor{black}{
The mismodelled $N-$body perturbations of the RA and DEC of Callisto are as large as $0.8$ arcsec and $0.6$ arcsec, respectively; they are about $3-4$ times larger than the LT shifts in  Figure\,\ref{fig1}.
}

\textcolor{black}{
Thus, improving the accuracy in the determination of the masses of the Galilean moons will be crucial for a successful test of the LT effect with the Jovian system.  The situation will become more favorable when, among other things, the masses of the three outer Galilean satellites will be accurately determined by JUICE and Clipper, while the flybys of Io by Juno should allow to improve also the mass of Io. More specifically, according to Tables 1 to 3 of \citet{2021AeMiS.100..195M}, the masses of Europa, Ganymede and Callisto should be determined by JUICE with an improvement of about $1-2$ orders of magnitude with respect to the errors retrieved at \url{https://ssd.jpl.nasa.gov/sats/phys_par/} and used in this paper.
}
\section{Discussion and conclusions}\lb{sect4}
The theoretically predicted LT centennial signatures of the Galilean moons of Jupiter seem to be large enough to be detectable, at least in principle, given the current level of accuracy in their astrometric measurements, of the order of $\simeq 10\,\mathrm{mas}=0.01\,\mathrm{arcsec}$. Moreover, also the systematic bias due to the multipolar gravity field of Jupiter appears to have a limited impact, especially in view of the likely improvements in knowing it after the completion of the ongoing Juno mission in the next few years. \textcolor{black}{On the other hand, the masses of the satellites need to be improved since the mutual $N-$body perturbations induced by one on each other are currently larger than the LT shifts, or represent a large fraction of the latter ones.}

It should be recalled that if a dynamical effect of interest, such as the LT one in the present case, is not explicitly included in the models fitting the observations, the resulting post-fit residuals may not display the sought signatures to a statistically significant non-zero level because they may partly be removed in estimating the initial state vectors and other parameters of the target system \citep{2023arXiv230301821F}. Despite a general rule valid in all cases cannot be given, it might be prudently said that to (partly) avoid to be absorbed in (some of) the estimated parameters of the fit, an unmodeled dynamical feature of motion should be much larger than the observational accuracy. Be that as it may, the correct procedure consists in proceeding on a case-by-case basis producing a new series of residuals constructed by explicitly modeling the effect of interest, determining some key parameters characterizing it, and inspecting the correlations among them and the other estimated parameters \citep{2023arXiv230301821F}. To the knowledge of the present  author, the gravitomagnetic field of Jupiter is currently included in the modeling of the orbits of Io, Europa, Ganymede and Callisto, but no dedicated data reductions aimed to explicitly detect the LT precessions of such natural bodies have been performed so far.

The present study suggests that the time to successfully do it may \textcolor{black}{come soon}, also in view of the expected improvements in  determining the orbits \textcolor{black}{and the masses} of  \textcolor{black}{Europa, Ganymede and Callisto} thanks to the forthcoming JUICE and Clipper missions. \textcolor{black}{Also the orbital and physical parameters of Io should be improved by its flybys by Juno, at least to a certain extent.}
\bibliography{Uranusbib}{}
\end{document}